*Review*

# Biodegradable Polymeric Micro/Nano-Structures with Intrinsic Antifouling/Antimicrobial Properties: Relevance in Damaged Skin and Other Biomedical Applications

**Mario Milazzo** [1,*] **, Giuseppe Gallone** [2] **, Elena Marcello** [3] **, Maria Donatella Mariniello** [4] **, Luca Bruschini** [5] **, Ipsita Roy** [6] **and Serena Danti** [1,2,4,*]

1   Department of Civil and Environmental Engineering, Massachusetts Institute of Technology, Cambridge, MA 02142, USA
2   Department of Civil and Industrial Engineering, University of Pisa, 56126 Pisa, Italy; giuseppe.gallone@unipi.it
3   School of Life Sciences, University of Westminster, London W1W 6UW, UK; elenamarcello@outlook.com
4   Doctoral School in Clinical and Translational Sciences, Department of Translational Research and New Technologies in Medicine and Surgery, University of Pisa, via Savi 10, 56126 Pisa, Italy; m.d.mariniello@gmail.com
5   Department of Surgical, Medical, Molecular Pathology and Emergency Medicine, University of Pisa, via Savi 10, 56126 Pisa, Italy; luca.bruschini@unipi.it
6   Department of Materials Science and Engineering, Faculty of Engineering, University of Sheffield, Sheffield S1 3JD, UK; i.roy@sheffield.ac.uk
*   Correspondence: milazzo@mit.edu (M.M.); serena.danti@unipi.it (S.D.)

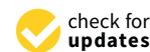

**Abstract:** Bacterial colonization of implanted biomedical devices is the main cause of healthcare-associated infections, estimated to be 8.8 million per year in Europe. Many infections originate from damaged skin, which lets microorganisms exploit injuries and surgical accesses as passageways to reach the implant site and inner organs. Therefore, an effective treatment of skin damage is highly desirable for the success of many biomaterial-related surgical procedures. Due to gained resistance to antibiotics, new antibacterial treatments are becoming vital to control nosocomial infections arising as surgical and post-surgical complications. Surface coatings can avoid biofouling and bacterial colonization thanks to biomaterial inherent properties (e.g., super hydrophobicity), specifically without using drugs, which may cause bacterial resistance. The focus of this review is to highlight the emerging role of degradable polymeric micro- and nano-structures that show intrinsic antifouling and antimicrobial properties, with a special outlook towards biomedical applications dealing with skin and skin damage. The intrinsic properties owned by the biomaterials encompass three main categories: (1) physical–mechanical, (2) chemical, and (3) electrostatic. Clinical relevance in ear prostheses and breast implants is reported. Collecting and discussing the updated outcomes in this field would help the development of better performing biomaterial-based antimicrobial strategies, which are useful to prevent infections.

## 1. Introduction

Bacterial colonization of implanted biomedical devices and prostheses (e.g., orthopedic prostheses, heart valves, breast implants, catheters, stents) is the main cause of healthcare-associated and nosocomial post-surgical infections, estimated to be 8.8 million per year in Europe. Some superbugs, like those encoding New Delhi metallo-beta-lactamase 1 (NDM-1), have sparked panic in several hospital areas, as they are resistant even to carbapenems, the antibiotics reserved as a last defense line [1]. Therefore, new antibacterial treatments are becoming vital to control nosocomial infections arising as surgical and post-surgical complications [2]. Many infections originate from damaged skin, which lets bacteria exploit injuries and surgical accesses as passageways to get to inner organs. Thanks to percutaneous implants and catheters used for draining, skin-resident microorganisms can migrate into the body and generate infections close to the implanted devices and/or in spare body tissues. Therefore, an effective treatment of skin damage is of utmost importance for the success rate of many biomaterial-related surgical procedures, and in general for healthcare and wellbeing [3].

Skin is the largest organ in the human body, and is the principal interface between the body and the surrounding environment. As such, it represents the first defense line against armful entities. Skin is considered as a bi-layered structure composed of epithelial (i.e., epidermal) and connective (i.e., dermal) tissues (Figure 1) [4]. The thin superficial layer of about 0.1 mm called the epidermis is made up of epithelial cells forming a multilayered squamous structure led by keratinocytes located above the basement membrane and corneocytes at the air interface, as well as including melanocytes and mechanoreceptors (i.e., Merkel cells) [5]. Beneath the epidermis is the dermis, made up by fibroblasts with their secretome (e.g., collagens, elastin), a dense network of innervation and vasculature, the latter providing the nutrients to the cells and appendages. Underneath, the subcutis (or hypodermis) is mainly composed of fat that acts as a buffer for both insulation and energy storage [6].

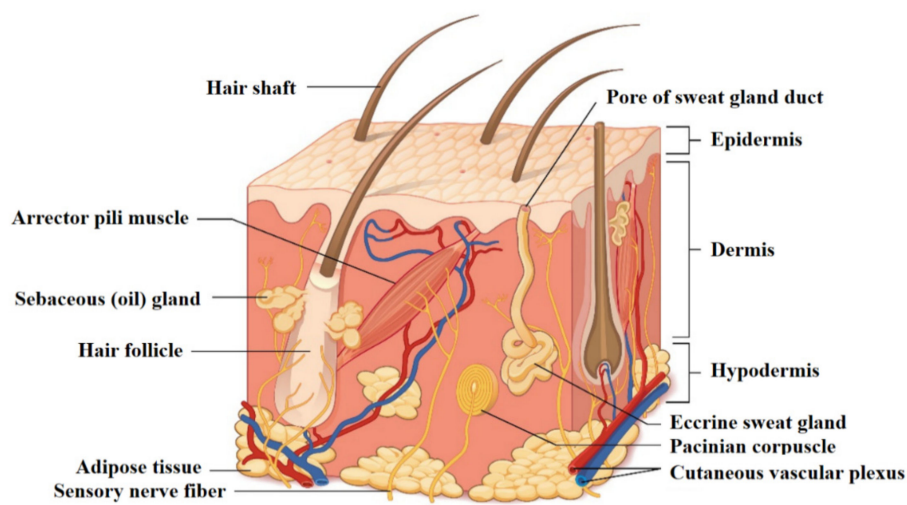

**Figure 1.** Structure of the skin: a superficial thin layer called the epidermis is the first barrier towards the external environment, provided with cells apt for protection and sensation. Beneath is the dermis layer, which provides both mechanical consistency and nourishing functions; it also hosts sensory receptors and vasculature. The underneath subcutaneous layer is mainly composed of fat for thermic insulation and energy storage. Reprinted with permission from [7] © 16 January 2020 OpenStax. Textbook content produced by OpenStax is licensed under a Creative Commons Attribution License 4.0 license.

Skin, the epidermis layer primarily, is permanently exposed to different chemical and physical factors that can potentially affect its structure or the tissues beneath it, causing the deterioration of the tissue and, potentially, the breakthrough of infections that can spread into the whole body [8]. Per se, human skin has a variegated microbiota, whose specific populations depend on the particular individual. The most common bacterial residents include *Staphylococcus*, *Corynebacterium*, *Cutibacterium*, *Micrococcus*, *Streptococcus*, *Brevibacterium*, *Acinetobacter*, and *Pseudomonas* [9]. In healthy conditions, the microbiota exerts a primary beneficial protective function; however, imbalances are associated with a number of skin diseases and infections. The most common and important class of skin diseases is caused by bacteria, either belonging to one's microbiota (including small intestinal barrier overgrowth) or coming from an external source. Bacterial-related pathologies can either be localized on the skin (e.g., cellulitis, impetigo) or affect other inner organs using the damaged skin as a passageway, and they emerge in a large number of diagnoses, being among the most frequent causes of infections in hospitalized patients [3]. Such clinical issues may even become enhanced when skin is damaged from burns, abrasions, or wounds [10]. Tissue engineering has developed different strategies to heal and replace the skin [11]. If the damage is limited to just the epidermis, no surgery is needed since self-regeneration processes take place thorough the keratinocytes still present in the site. In case the number of cells is not sufficient, it is possible to empower the regeneration process by the use of epithelial stem cells, e.g., from follicles. In contrast, if the defect extends deeper in the skin, homologous/synthetic grafts are employed on injured areas, possibly with surgical intervention to place them and close the wound [12].

The healing process is a delicate procedure, during which there is a high risk of infection since the damaged skin is not able to oppose a proper barrier to chemicals or bacteria. This is the motivation for using antibiotics to cure infections that have greatly reduced the mortality from bacterial pathologies worldwide. Unfortunately, the mutation of bacteria, leading them to become antibiotic-resistant, has significantly changed the curing trend [13,14]. Alternative approaches to antibiotics are nowadays represented by antimicrobial and antifouling solutions (Figure 2).

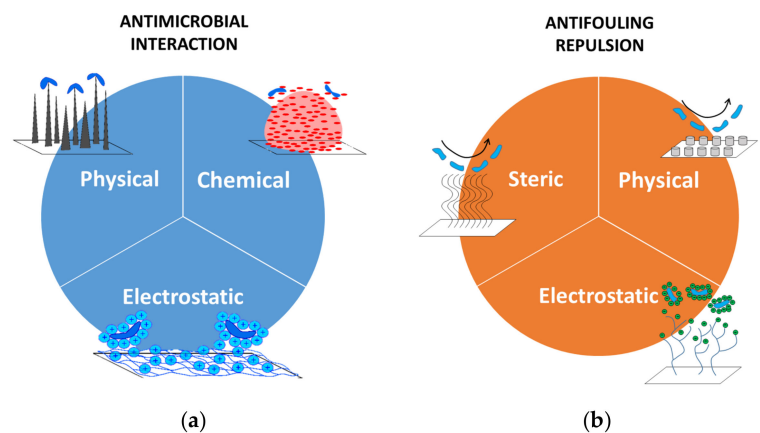

**Figure 2.** Antimicrobial and antifouling strategies. (**a**) Antimicrobial solutions aim at killing living microorganisms through physical (mechanical), chemical, or electrostatic interactions. (**b**) Antifouling systems contrast living organisms or inert bodies (e.g., dirt) through steric, physical (morphological), or electrostatic repulsive actions. Parts of the figure are reprinted from [15]—Open Access under Creative Commons CC BY 4.0 license.

The first class includes systems aimed at killing undesired living organisms by means of either chemical interactions (e.g., drug release [16]), physical contact through a modification of the nanostructure, or electrostatic neutralization of the activity of the living microbes. Once microbes have been killed, they lose their harmful potential and are promptly removed by the immunity system that, thanks to phagocytes, prevent the permanence of such biological structures. Antifouling solutions have a broader impact because they prevent the adhesion on a surface of microorganisms, but also inert bodies (e.g., dust). Their working principles are slightly different from those of the antimicrobials', and are based on the triggering a repulsion activity of some kind. A first case is the steric repulsion, in which adsorption is prevented by the compression of polymeric chains that cover a surface and behave as mechanical springs. A second class includes the modification of the surface through patterns that establish morphological conditions unfavorable to the attachment of foreign bodies. This is the case, for instance, of the so-called superhydrophobic structures [17]. The last two approaches consist in exploiting physical properties of surfaces that are able to promote repulsion forces, such as, for example, promoting a low surface energy or inducing an electrostatic polarization.

Antimicrobial and antifouling solutions possess a great appeal since they do not require the use of drugs, and thus they are likely to be healthier for the patient, and also they are to a large extent insensitive to mutation of the targeted bacteria. Starting from such considerations, in our review we focus on the design and fabrication of biodegradable structures that exhibit antimicrobial and antifouling properties without releasing additional substances.

Usually, most of the microorganisms that make up the skin flora establish a mutually beneficial relationship with the body. In fact, they hinder the colonization of pathogens by subtracting their nourishment by producing antimicrobial substances and by lowering the skin pH via the degradation of the sebum they feed on. Others, such as *Staphylococcus aureus* or *Candida albicans*, although potentially pathogenic, do not form numerically sufficient colonies to cause problems for a healthy organism.

Nevertheless, under some circumstances, protein absorption and bacterial adhesion can boost the formation of cell clots and biofilms, which can lead to implant rejection or failure, up to death if they migrate to vital organs or become resistant. Bacterial colonization phenomenon at the implant site is enhanced by the formation of a non-vascularized fibrotic scar around the device and is known as biofouling. Once a site becomes infected, an antibiotic therapy is applied. However, since only a low dose can reach the infected area across the avascular fibrotic scar, poor efficacy is observed after a first healing, which in time is followed by the development of chemically resistant species and consequent microbial reinforcement. Infections with *S. aureus* are in fact the prevalent device-related infections. Recently, infections associated with heart valves contaminated by nontuberculous mycobacteria (NTM) highlighted the hydrophobic ability of these cells, which preferentially attach to surfaces to form thick biofilms containing high numbers of cells, and thus the layers of cells and extracellular materials in biofilms substantially develop disinfectant-resistance.

Surface coatings can avoid biofouling by using appropriate physical and morphological properties of materials, thus inhibiting bacterial colonization by means of non-chemical mechanisms, which therefore do not induce bacterial resistance. By providing unfavorable surfaces for bacterial adhesion and/or presenting physical barriers at relevant orders of magnitude to isolate single cells or cell groups, reduced numbers of bacteria are finally present with a difficulty to grow and generate biofilms, thus becoming more vulnerable to the immune system. In this respect, super-hydrophobicity is a highly desired property to limit the biofouling phenomenon. A super-hydrophobic surface prevents wettability (therefore, protein absorption), forming single water drops with a static contact angle >150° with the surface. The physico-chemical properties of superhydrophobic materials have been a subject of study in the last decade due to the relevant implications in industry and healthcare sectors. Nanostructured surfaces, like pillar or sharklet-like patterns, offer valuable anti-biofouling properties [17]. However, some issues have to be solved to obtain industrial-scale applications, including costs and durability of micro/nanostructured surfaces [18]. Recently, electrospinning has been used to fabricate superhydrophobic materials, as the small fiber diameter has shown to contribute to super-hydrophobicity by the surface roughness/texture resulting from the superposition of fibrous layers [19]. Electrospinning is a cheaper and up-scalable technique that, when compared to other nanopatterning systems (e.g., etching, photolithography), results in being more effectively applicable to industrial-scale manufacturing of devices. However, many parameters (voltage, feed rate, viscosity, electrical properties of the solution, set-up geometry, and environmental conditions) affect the final result and must be optimized in order to obtain the desired fiber size, interspace/porosity, alignment, surface roughness, and other specific properties, which greatly enhance the difficulties in setting up the process. Among the different biomaterials reported with antimicrobial properties, biodegradable polymers offer interesting features in biomedical applications. A scientifically recognized definition for "biodegradable polymer" is "a material for which the degradation is mediated, at least, partly from a biological system" [20]. The disappearance of a biomaterial over time after implantation has been related to three distinct phenomena: biodegradation, bioresorption, and bioabsorption. Biodegradation is intended when the polymer chain breaks into natural byproducts processed by biological agents present in the microenvironment, such as enzymes, leading to material disintegration, erosion, or dissolution. Bioresorption occurs when the degradation products of polymers are resorbed in the body by a metabolic process. Instead, bioabsorption involves polymers that dissolve in biofluids and are eliminated without chain scission, such as poly(vinyl alcohol) (PVA) and poly(ethylene glycol) (PEG).

The focus of this review is to highlight the potential role of degradable polymeric micro- and nano-structures (most of all biodegradable, but including other degradable polymers) that show intrinsic antifouling and antimicrobial properties, with a special outlook towards biomedical applications dealing with skin and skin damage. Collecting and discussing the updated outcomes in this field would pave the development of better performing biomaterial-based antimicrobial strategies, useful to prevent and control surgical and post-surgical infections. As clinically relevant examples, this review shows usefulness in the complications of breast implants and otorhinolaryngology prostheses, such as tympanic membrane and nasal septum repair devices. The most widely used breast implants are gel-filled silicone shells, categorized into different surface types according to surface roughness and dimensionality ratio. The most common complication with a 10.6% overall incidence is the capsular contraction, which is foreign body reaction. As the scar is thick and avascular, microbes migrating from catheters to the implant surface are very difficult to eradicate with systemic antibiotic therapy and exacerbate the inflammatory process. Ear and nose are easily contaminated and stressed by bacteria, for example in cases of recurrent otitis and rhinitis. Many types of prosthetic devices are available to repair ear and nose damaged tissues; however, the two most common end-fates for synthetic and biological materials in such infected sites are extrusion and resorption, respectively, which are both driven by inflammatory processes [21,22]. Overall, new approaches to reduce bacterial contamination of surfaces would enable better-performing biomedical devices to treat skin damage and prevent post-surgical complications in many medical areas.

## 2. Antimicrobial Approaches

Antimicrobial approaches to treat damaged skin aim to kill microbes by exploiting either peculiar surface topologies or the physico-chemical properties of cationic polymers. Although the chemical approach represented by antibiotic-loaded hydrogels have been a reliable approach to dress wounds due to their antimicrobial properties [16,23], we have decided to exclude this topic since our focus concerns materials/topologies with intrinsic antimicrobial features. Before going into detail of each class, we have displayed Table 1 below, which summarizes the main materials/features later discussed.

### 2.1. Surface Topology

The antibacterial effect based on mechanical interactions was originally observed in nature [59]. For instance, wings of insects such as cicada or dragonflies possess a nanopatterned surface with high-aspect-ratio cone-like nanopillars that is lethal for bacteria such as *Pseudomonas aeruginosa*. It has been speculated that this is the effect of an evolutionary adaptation to the environment, preventing the formation of biofilms that affect the aerodynamicity of such insects [24,26–28].

The antimicrobial mechanisms of nano-structured/patterned surfaces have been deeply investigated by studying the influence of specific parameters, such as the distribution density and the shape of nanopillars. Scientists have shared different opinions—from one side, Li et al. stated that an increased number of nanopillars per surface unit is beneficial to promote a larger adhesion area and, thus, the lethal effect [29]. This result was also confirmed by Linklater et al., as well as by Kelleher et al., who assessed the performance of both synthetic nanostructures made of silicon and of cicada wings [24,30]. In contrast, Xue et al. [33] theorized that a lower density associated with the use of sharp nanopillar could be more effective, and later Fisher et al. [31] experimentally demonstrated this using *P. aeruginosa*.

**Table 1.** Summary of the antimicrobial materials/features reported in this work.

| Classification | Material/Feature | Advantages | Disadvantages | Reference |
|---|---|---|---|---|
| Surface topology | Made-up nanopillars<br>Diamond<br>Gold<br>Polymethyl methacrylate (PMMA)<br>Titanium | Independent of the material properties | Made-up structures are potentially costly in terms of manufacturing | [24–33]<br>[31]<br>[34]<br>[35]<br>[36,37] |
| Cationic polymers | Polymers with ammonium ions<br>Polymers with sulfonium ions<br>Polymers with phosphonium ions<br>α–helical structures<br>Chitin<br>Chitin nanofibrils (CN)<br>Polymer/CN composites<br>Polyhydroxyalkanoates (PHAs) with thioether/<br>thioester moieties (PHACOS) | No additional fabrication costs to implement antimicrobial properties (intrinsic features) | Dependent on the physico-chemical properties of the materials | [38–42]<br>[43]<br>[44,45]<br>[46–50]<br>[51–53]<br>[54,55]<br>[56]<br>[57,58] |

However, scientists agree on two aspects: antimicrobial efficacy is improved if the nanopillars have a high aspect ratio (e.g., diameter of 20–80 nm and height of 500 nm [32]) and if the bacteria have a high motility, since it promotes a larger contact [24,25]. Bacteria are, in fact, not killed by a sudden interaction, but by the prolonged contact that leads to the elongation of the cellular structures and, ultimately, to unbearably high shear stresses (Figure 3) [26].

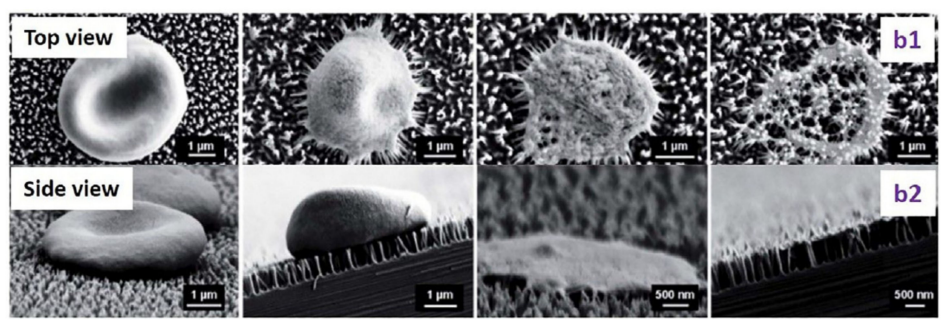

**Figure 3.** SEM images of the interaction of an erythrocytes with nano-pillared surfaces. Top (**b1**) and side (**b2**) views of the morphological changes of the erythrocyte. Reprinted with permission [60] © 2014 Royal Society of Chemistry.

A large number of antimicrobial surfaces have been developed by taking inspiration from natural systems and employing different materials (e.g., diamond [31], gold [34], polymethyl methacrylate (PMMA) [35]), but only silicon-based structures have demonstrated high in vitro and in vivo biocompatibility [61–63]. Studies in mice have demonstrated how the same material, but with a different patterning, is able to reduce inflammations after surgery [61]. Another interesting material is titanium, which has been used in many prosthetic applications as a substitute of the native cortical bone (e.g., [36,37]). However, the antimicrobial properties of titanium-based structures have been proven only in vitro, and an in vivo assessment is still missing [64–66].

## 2.2. Antimicrobial Cationic Polymers

Antimicrobial cationic polymers are usually employed in devices for clinical applications, ranging from medical devices to wound treatments [67–69]. This class of polymers base their antimicrobial action through two main functional components: cationic and hydrophobic groups. The first class allows adsorption onto the membranes of the microbes while the hydrophobic components cause cytoplasm leakage and the ultimate death of the pathogenic cells [70,71]. Given this peculiar mechanism, the structural design of such polymers assumes a key role in determining and tuning the antimicrobial properties [72]. The primary structure of antimicrobial cationic polymers concerns cationic groups that are usually ammonium [38–42], sulfonium [43], and phosphonium ions [44,45]. As for the hydrophobic groups, the most used are alkyl chains whose length may or not trigger the lethal effects on microbes [73]. In fact, non-optimized long chains can weaken the biocidal activity and, at the same time, increase the hemolytic activity [39,74]. The other important aspect is the spatial arrangement of the polymer molecules, whose chains are mainly organized in α-helices [75]. A correct control of such a secondary structure can, indeed, boost the antimicrobial properties of the polymer by properly positioning the cationic and hydrophobic functional groups along the molecular chain [76]. In this respect, very effective α-helical structures have been found, which, due to the organization of their hydrophilic and hydrophobic groups, are able to promote a cationic antimicrobial effect [46–50]—hydrophilic regions with positive charges are absorbed by the microbe while the hydrophobic ones cause the death of the pathogens [77,78]. Many studies have been carried out to mimic the behavior observed in nature [79–82]; for instance, Gellman et al. synthetized an artificial α-helical polymer that was effective against *Escherichia coli*, *Bacillus subtilis*, *Staphylococcus aureus*, and *Enterococcus faecium* [83].

Chitin is a natural polysaccharide found particularly in the shells of crustaceans, cuticles of insects, and cell walls of fungi. A deacetylated form of chitin with deacetylation greater than 50% is chitosan, a well-known biopolymer that has been reported to have antibacterial properties [51–53]. Chitosan is metabolized by some human enzymes, such as lysozyme, and thus it is considered bioresorbable. Owing to the large quantities of amino groups on its chain, chitosan dissolves at low pH (i.e., acidic solutions), showing a pH-sensitive behavior as a weak polybase due to the protonization of swollen amine groups under low pH conditions. Due to its pro-inflammatory activity, chitosan plays a role in the wound healing process. Since blood cells have a negatively charged surface, contact with chitosan adheres tightly to the wound and stops bleeding.

Differently from chitin, which exhibits immunogenic properties, and chitosan, which is pro-inflammatory, nano-sized (<0.2 μm) chitin shows anti-inflammatory along with antibacterial properties [54,55]. Chitin nanofibrils (CN) represent the purest crystal form of chitin and show positive surface charges due to the amino groups (Figure 4). In skin contact application, CN exhibited very good anti-inflammatory properties; for example, upon in vitro administration to human keratinocytes (HaCaT cells) at 10 μg/mL, CN reduced many pro-inflammatory interleukins (IL)—IL-1α, IL-1β, IL-6, and IL-8—and tumor necrosis factor alpha (TNF-α) in 6–24 h [84]. CN also showed good direct and indirect antimicrobial properties, the latter through stimulating the innate immune response of skin cells [85,86]. One peptide involved in skin-mediated immunity is the human beta defensin 2 (HBD-2), which acts as an endogenous antibiotic against Gram-positive and Gram-negative bacteria, fungi, and the envelope of some viruses [87]. Danti et al. [84] reported that CN–nanolignin microcomplexes administrated in culture at 0.2 μg/mL to human dermal keratinocytes (HaCaT cells) were able to increase HBD-2 expression of about 140% with respect to basal conditions, which can be relevant in skin self-defense. The indirect antimicrobial properties of CN could also be found in biodegradable polymeric nanocomposites proposed for skin contact applications. However, a high quantity of CN is needed on the surface to obtain a direct antimicrobial effect, which is difficult to achieve using polymer/CN composites [56]. Therefore, CN surface coatings seem to be more suitable for inducing antimicrobial activity than CN bulk incorporation. In bionanocomposites, CN can be released upon polymer degradation to sustain anti-inflammatory activity. CNs can be metabolized from the many families of chitotriosidases present into the human body.

Sulfur has been exploited as an antimicrobial agent since ancient times. Nevertheless, limited research has been conducted on antimicrobial polymers incorporating this element. Alongside the aforementioned sulfonium groups, recently a class of natural polymers containing sulfur, i.e., polyhydroxyalkanoates (PHAs) with thioester moieties (i.e., PHACOS), have been proven to exhibit antibacterial properties [57]. PHAs are bacteria-derived polyesters, an attractive class of biocompatible materials for a range of medical applications, including skin [88,89]. They are bioresorbable and biobased polymers that can be obtained from renewable resources, making them a valuable alternative to synthetic petroleum-derived plastics [90]. PHACOS are PHAs with thioester linkages in their side chains, possessing antimicrobial activity selectively against *S. aureus*. The presence of the thioester group has been associated with the biocidal activity of these polymers [57,58]. Dinjaski et al. showed that a direct contact between the bacteria and the material was essential for the antimicrobial activity of PHACOS, suggesting a possible interaction between the thioester groups and the bacterial membrane [57,58]. The alkyl chains present in the structure of PHACOS (Figure 5) could also participate in the biocidal activity of the material, as described in the previous section.

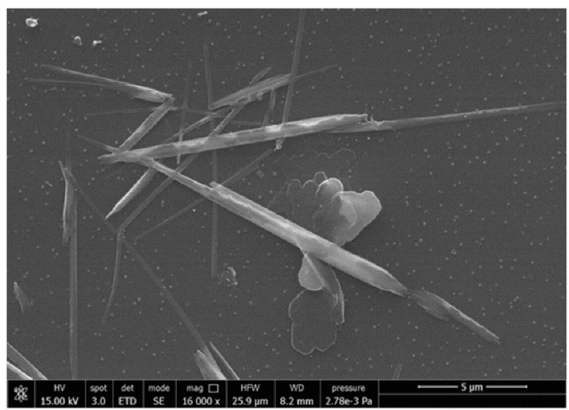

**Figure 4.** Scanning electron microscopy (SEM) image showing CN. Reprinted from [84]—Open Access under Creative Commons CC BY 4.0 license.

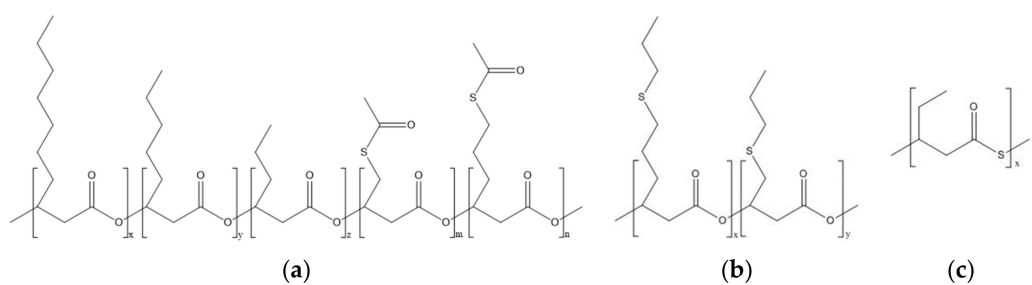

**Figure 5.** Chemical structure of polyhydroxyalkanoates (PHAs) containing sulfur. (**a**) PHAs with thioester linkages in their side chains (PHACOS). (**b**) PHAs with thioether linkages in their side chains (poly(3HPTB-co-3HPTHx)). (**c**) PHAs with thioester groups in their main chains (poly(3MV)).

Nevertheless, the exact mode of action of PHACOS and thioester groups has not yet been elucidated. To date, PHACOS are the first and only PHAs to show inherent antibacterial properties. Other sulfur-containing PHAs have been produced, containing thioester groups in their main chains [91] or thioether linkages in their side chains [92], but their possible antimicrobial activity has not yet been analyzed.

Another interesting class of polymers containing sulfate groups is polysaccharides from seaweeds. Different subclasses of sulfated polysaccharides have been identified according to the algal origin: ulvan from green seaweeds, fucoidan from brown seaweeds, and galactan and carragenan from red seaweeds [93]. The sulfate groups, and in particular their position in the polymer chain, were found to be responsible for a set of bioactive properties, encompassing antioxidant, anti-inflammatory, and antiviral activities [94,95].

## 3. Antifouling Approaches

In contrast to antimicrobial approaches, antifouling materials/features (summarized in Table 2) prevent the adhesion of microbes/external bodies. The following subsections aim to describe the main applications for treating damaged skin.

**Table 2.** Summary of the antifouling materials/features reported in this work.

| Classification | Material/Feature | Advantages | Disadvantages | Reference |
|---|---|---|---|---|
| Steric | Polymer zwitterions | No additional fabrication costs to implement antifouling properties (intrinsic features) | Dependent on the physico-chemical properties of the materials | [96–102] |
| | Ulvan | | | [103–107] |
| Surface topology | Superhydrophobic surfaces | Independent of the material properties | Potentially costly in terms of manufacturing | [108–122] |

### 3.1. Steric Repulsion

The mechanism behind steric repulsion involves long-chain polymers stabilized on a surface, which prevent bacteria adsorption. Among the most common polymers used for this purpose, PEG and poly(ethylene oxide) (PEO) are two polyethers that possess the same repeating unit but with different molecular weight. PEG/PEO are important materials for producing antifouling surfaces since they have a low surface energy (below 5 mJ·m$^{-2}$) and make weak bonds with proteins [123]. These polymers are bioabsorbable materials, and thus they are eliminated by the excretory organs without chain scission. As stand-alone, they are not advisable for human body-related applications, since they can trigger anaphylactic reactions and hypersensitivity reactions [124,125]. PEG and PEO are thus used to obtain polymer-based composites as dispensing agents for the filler, such as CN/Polylactic acid (PLA) composites [56,126].

Polymer zwitterions are a second class of materials that are exploitable for antifouling purposes thanks to their low energy and structure [99,100]. They are chemically stable and are able to pull water and foreign organisms away, creating a barrier, and thus being more suitable for tissue engineering applications [101,102]. Polymer zwitterions are usually electrospun from precursor solutions with low concentration and viscosity [97,98]. Remarkable applications include the fabrication of vascular grafts using polyurethanic matrices [96].

The antimicrobial activity of ulvan has been recently reported [105,107]. Ulvan shows good biocompatibility and antifouling properties, thus appearing as a green material with prospects in functional coating biomedical implants and devices. Sulfate groups specifically impart ulvan with antiviral activity against herpes simplex virus 1 (HSV-1) and paramyxoviridae [103,104]. By means of its antiadhesive character led by its negative charges, and independently of surface roughness, ulvan was able to inhibit biofilm formation and bacterial adhesion by *P. aeruginosa* on titanium surfaces [106].

### 3.2. Surface Topology

As stated and described in Section 2.1., the antifouling effect based on surface topology also has its roots in the observation of nature. Scientists have observed and studied a large number of natural systems that use this feature as a mean to adapt to the surrounding environment and to enable specific features such as protection against external organisms or improvement of dynamic or self-cleaning performances. Scientists have identified different mechanisms under the umbrella of micro-topology modification. The most important, and common, is the superhydrophobicity—a surface is defined as superhydrophobic if the contact angle generated by the deposition of a water droplet is between 90° and 150°. Without a perfect attachment, water can remove contaminants by rolling off the surface [110,115]. Similar effects have been investigated at different scales to prevent the attachment of cellular tissues or macro-organisms while dynamically exposed to drag forces [112–114]. In contrast, a completely different mechanism is represented by animal secretions that dynamically influences the antifouling properties of the surfaces [111,122].

Natural antifouling surfaces exploiting topological features have been found in about 900 marine animals of the Elasmobranchii family [121]. This group includes sharks, skates, and rays, whose placoid scales are composed of a vascular core of dentine beneath an acellular layer of enamel [119,120]. These structures have a undulated topology that enables a passive cleaning induced by drag forces and, at the same time, improves the hydrodynamicity of the animal. On the basis of the species, the specific topology of such structures, and thus the associated features, slightly change for the best adaptation to the surrounding environment (Figure 6a,b).

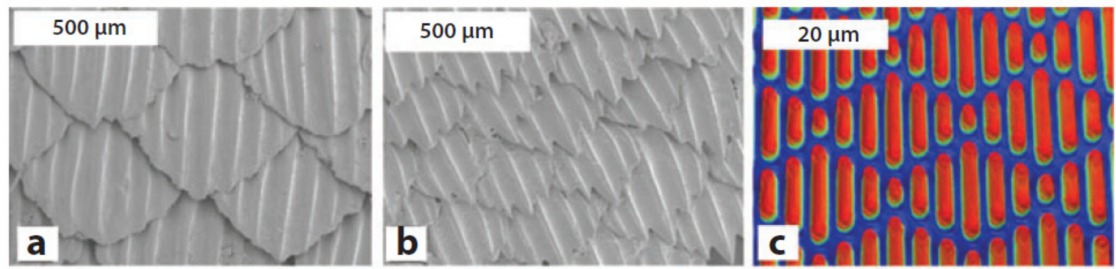

**Figure 6.** SEM micrographs of (**a**) spinner shark skin and (**b**) Galapagos shark skin surfaces. (**c**) Optical profilometry of sharklet surface. Reprinted with permission [127] © 2010 Elsevier.

The self-cleaning mechanisms activated by superhydrophobicity have been observed on lotus and rice leaves, above which water drops are able to move freely, collecting foreign living and inert bodies and transporting them passively thanks to inertial forces. The so-called lotus effect is due to the microstructure of the leaf surface that presents a pattern of shaped cones with nano-scale hair-like structures, which induces the superhydrophobic effect (Figure 7a) [116–118].

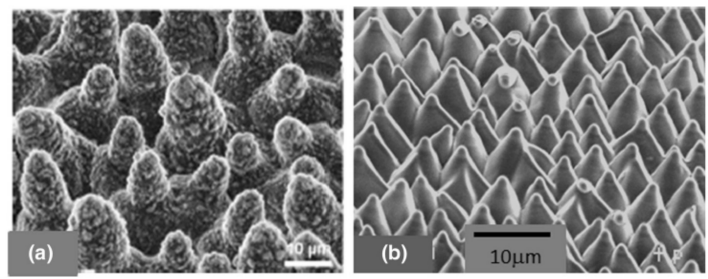

**Figure 7.** (**a**) SEM images of the micro-surface of a leaf of a lotus plant with the characteristic conical cell pattern. Reprinted from [118]—Open Access under Creative Commons CC BY 4.0 license. (**b**) Bioinspired cone-shaped patterned surface resembling the microstructure of the lotus leaf. Reprinted with permission [109] © 2006 American Institute of Physics.

Artificial attempts to shape surfaces with bioinspired antifouling properties have been carried out using lithography and self-assembly techniques [108,109,115]. Following this approach, a number of synthetic products have been developed and commercialized, exploiting the peculiar antifouling characteristics. Remarkable examples are the devices made in either polydimethylsiloxane or thermoplastic polyurethane polymers by Sharklet Technologies, whose micropatterned surfaces aimed at mimicking shark skin (Figure 6c) or lotus micro-surface (Figure 7b) were effective in preventing bacterial adhesion with direct implications on skin-related applications [128] or laser-ablated films.

## 4. Clinical Relevance and Future Perspectives

We present a review on emerging approaches to prevent skin and skin-derived infections using biodegradable polymeric micro/nano-structures with intrinsic antifouling/antimicrobial properties as a valuable alternative to traditional pharmacological antibiotic treatments or drug release materials [16] that can generate resistant bacteria. Degradable biomaterials, such as properly biodegradable as well as bioresorbable and bioabsorbable, play a key role in a number of biomedical applications, including, but not limited to, skin repair [23]. Moreover, the cosmetic and sanitary industries offer a set of skin care products and procedures to clean, restore, reinforce, protect, and maintain skin wellbeing, which can all benefit from the possibility of antifouling and antimicrobial surfaces. The functionalities expected by sanitary and cosmetic products are less strict than those required by biomedical devices, however, they may require materials/additives to treat potentially contaminated skin and bacterial proliferation (e.g., diapers); allow skin cleaning from dirt, sebum, microorganisms, and dead cells (e.g., beauty masks); reduce inflammation; and allow healing.

Depending on the extent of damage and presence of systemic pathologies (e.g., diabetes), skin wounds caused by burns, soars, or cuts are exposed to microbial contamination for several weeks and need to be properly treated. The physiological process that leads to the repair of a wound relies on the activation of a series of specific events involving a large number of cells and molecules. The process can be divided into three phases that overlap each other (Figure 8): (a) inflammation, (b) tissue formation, and (c) tissue remodeling. Some types of wounds, such as diabetic foot ulcers, venous ulcers, and some surgical wounds fail to complete this chain of events, thus becoming chronic wounds. These types of wounds often remain stationary at the first stage [129] and show a high presence of proinflammatory cytokines, high protease levels, many neutrophils, and senescent cells that do not respond to stimuli [129,130]. The wound bed can remain open for months or years, leading to repeated bacterial infections [131]. It is a fact that damaged skin offers a passageway for skin-resident and nosocomial bacteria to migrate to inner organs due to continuity between sterile and non-sterile compartments of the body. In skin dressings used to treat chronic wounds, a weak antimicrobial activity combined with defensin (e.g., HBD-2) expression by keratinocytes is desirable to maintain a safe microenvironment and allow the skin self-repair process.

## Wound healing

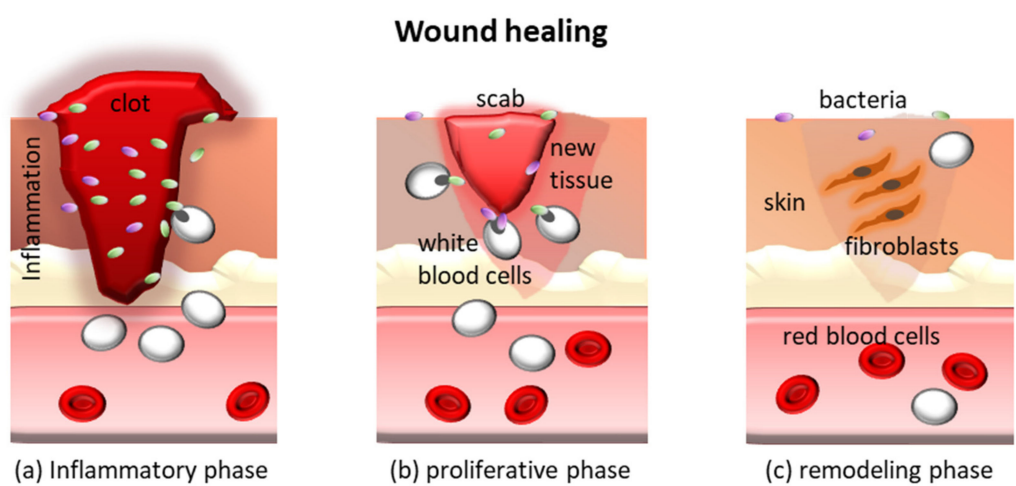

**Figure 8.** Schematic of the three phases of the wound healing process.

Chronic wounds are indeed very difficult to treat, representing sites at risk of infections since they are usually unhealed for more than 3 months [129]. In these cases, often secondary to other pathologies such as diabetes that decrease the skin regenerative capacity, locally aggressive antimicrobial treatments, including those containing silver, can further hamper the wound repair, whereas long-term antibiotic treatments are not recommended [4]. Therefore, biodegradable polymers with intrinsic antifouling/antimicrobial activity are highly suitable to produce wound dressings. One expected action of these materials is to combine three key properties, which all concur to the optimal wound healing cascade: (a) immunomodulation (including indirect antibacterial effects), thus controlling the inflammatory phase; (b) biocompatibility combined with antimicrobial properties, thus allowing the optimal timeline of tissue formation phase; and (c) degradability to allow space for tissue growth and remodeling. Since the cells involved in wound healing process, i.e., keratinocytes, fibroblasts, Langerhans cells, and recruited immune cells, concur to the eradication of bacteria, properly designed bio (nano)-materials can drive the healing process by regulating the immune cell trafficking and cross-talk at the wound site [132,133].

Many body parts have been reported as targets of infections and, in particular, of bacterial biofilms, as a consequence of device implantation [134] (Figure 9). In fact, permanent implants and endoprostheses (e.g., breast implants, orthopedic prostheses, implantable ear devices, and many others) pose the problem of surgical and post-surgical contaminations. In these cases, the stronger the immune reaction, the thicker the avascular fibrotic layer and the less efficient the effect of systemic antimicrobial treatments [135]. Therefore, bacteria that have reached the implant surface can proliferate undisturbed. Under these circumstances, intrinsically antimicrobial degradable materials could be used as surface coatings with a primary role in controlling immune reactions and bacterial infiltration in the initial stage post-implantation, namely, when fibrotic encapsulation takes place. Emerging robotic implants also raise the question of controlling undesired reactive phenomena such as contaminations and inflammatory reactions, which ultimately affect biocompatibility [136].

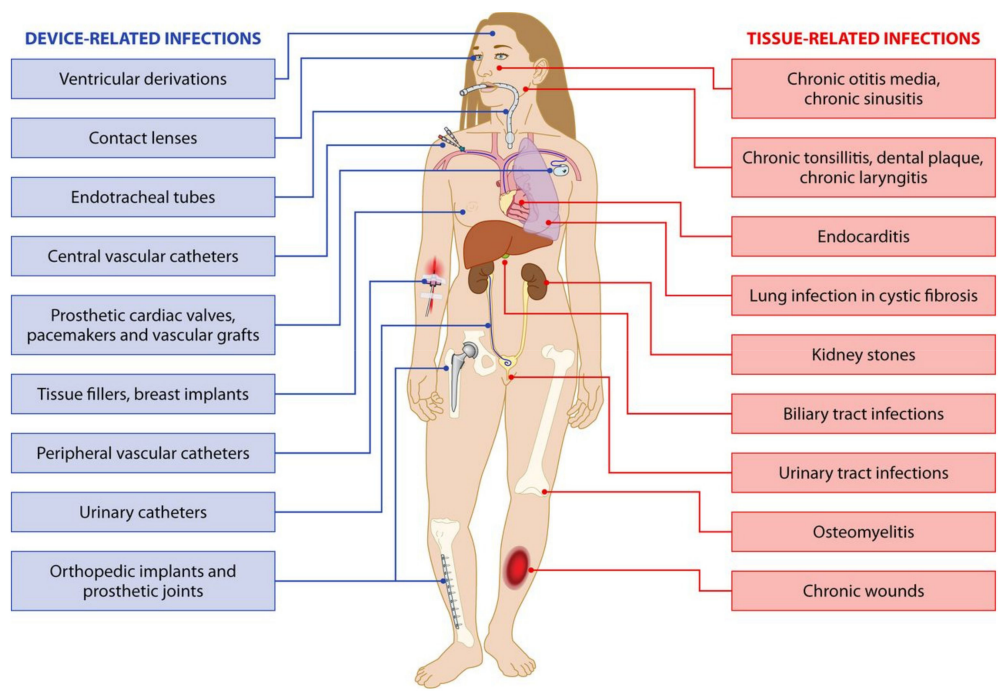

**Figure 9.** Reported sites of device-related and tissue infections. Reprinted with permission [134] © 2014 American Society for Microbiology.

Depending on the specific clinical issue, the antifouling/antimicrobial features of such structures can be exploited either as surface coatings or as actual structural medical devices. Usually, different biomedical devices and implants have been reported to show relevant incidence of infection [137]: 16–36% of left ventricular assist devices, 5–20% of cardiac implantable electronic devices, catheters (20% of peritoneal, 4.2–5.3% of dialysis, 2.5–4% of central venous, 1–3.4% of arterial, and 3.1–7.5% of urinary), 1–10% of meshes for ventral hernial repair, and 5% of elbow joint prostheses. Catheters, in particular, play a main role in infections due to the permanence of a percutaneous access to skin and environmental bacteria. In this review, we report here specifically on other settings dealing with complications for bacterial infections, namely, breast implants and otorhinolaryngologic implantable devices.

Breast implant infections are the leading cause of morbidity after breast surgery, with an incidence rate of 1% to 2.5% [138]. The incidence is higher after surgery for breast cancer (especially mastectomy followed by radiotherapy) than after breast augmentation and surgical technique, and patient's conditions are considered the most important determinants. In most cases, this complication develops within the acute post-operative period, but it is reported also years after surgery. In the majority of cases, it is caused by Gram-positive pathogens, such as *Staphylococcus aureus*, and empiric initial treatment is based on the prevalence of methicillin-resistant staphylococci [138]. Determining the origin of implant infections is still a challenge but potential sources of early post-operative infections include a contaminated prosthesis, the surgical environment, the patient's mammary ducts or skin, and the prosthesis contamination from remote infection sites. On the contrary, late infections usually are consequent to secondary bacteremia or an invasive procedure in other body districts [139]. Depending on severity of the infection, patients are treated with empiric intravenous or oral antibiotics and closely monitored [138]. All attempts should be made to salvage the implant. In cases of failure of the medical therapy, it is necessary to combine implant removal and antibiotic treatment followed by delayed reconstruction with a new operation weeks/months after the onset of the complication [138]. This approach requires multiple surgical procedures with increased risks and costs. Furthermore, delayed implant re-insertion may be technically more difficult due to tissue fibrosis. Breast implant infections are overall responsible for 2.0–2.5% of re-interventions performed in breast surgery [139] and low-grade or subclinical infections probably have a role also in the origin of capsular contracture. Therefore, new materials to coat breast prostheses able to reduce contaminations are expected to improve the success rate of these prostheses.

Antifouling/antimicrobial coating seems to be useful for a number of biomedical devices including the auricular implants. The aim is avoiding the infections and the consecutive extrusion of the implanted devices, which is a frequent event due to frequent otitis affecting these patients. Preliminary testing of a thin antifouling coating for a new type of cochlear implant have been evaluated [140]. Similar materials could be utilised for middle ear implants such as bone-anchored hearing aids. These devices often cause a skin reaction, and an antifouling/antimicrobial coating could solve this problem.

Moreover, these materials could find an application as surgical devices in ear, nose, and throat (ENT) procedures. At the end of the ear and nose surgery, de-epithelialized areas of the external or middle ear or the nasal septum or fossa have to be protected with silicone sheets or other materials to avoid scar formation. A layer of antifouling/antimicrobial material could improve the protection and facilitate the skin restoration. Equally, thin sheets coated with antifouling/antimicrobial biomaterials show potential to seal eardrum or nasal septum perforations. The chronic perforation of the tympanic membrane or of the nasal septum needs a difficult surgical reparation and the new antifouling/antimicrobial materials could easily close the hole. In particular, for nasal septum perforation, a hole greater than 2 cm significantly increases the surgical failure rate [141]. In case of surgical failure, a nasal button available on the market can be a solution. However, as a foreign body, this material can promote the formation of mucous scabs, creating an obstruction to the air flow. On the contrary, a nasal button covered with an antifouling/antimicrobial coating could seal the septal perforation, avoiding the surgical procedure and the formation of mucous scabs, thus restoring tissue function.

Proper material selection, design, and manufacturing of the surfaces of biomedical devices and implants, in particular considering their physico-mechanical, chemical, and electrostatic characters, are of paramount importance to reduce infection-related complications, thus allowing a better performance of the implanted devices. In light of this, additive manufacturing technologies may represent the key to synergistically exploit topologies and biomaterials as a powerful alternative to the traditional methodologies (e.g., etching), creating optimized surface topologies made of tuned materials able to either kill or repulse microbes. Examples of such an approach are the development of optimized antimicrobial/antifouling fibers fabricated via electrospinning [15,142], tuned hydrogels fabricated with 3D printing technologies [16,23], or the employment of stereolithography to modify hydroxyapatite-based composites for manufacturing antimicrobial dental bites [143].

The process by which bacteria adhere to, colonize, and infect a device is not fully understood, as it occurs on a multi-length scale and involves several physical, chemical, and biological aspects [137]. Therefore, biomaterial science integrated with nanotechnology and microfabrication could empower the intrinsic ability of some natural and biosynthetic polymers towards safer and better performing biomedical devices.

**Author Contributions:** Conceptualization, M.M. and S.D.; methodology, M.M. G.G., and S.D.; investigation, M.M. and S.D.; writing—original draft preparation, M.M., E.M., M.D.M., L.B., and S.D.; writing—review and editing, M.M., G.G., and S.D.; visualization, M.M.; supervision, G.G. and S.D.; project administration, I.R. and S.D.; funding acquisition, I.R. and S.D. All authors have read and agreed to the published version of the manuscript.

**Funding:** The project received funding from the Bio-Based Industries Joint Undertaking (BBI JU) under the European Union's Horizon 2020 research and innovation program by POLYBIOSKIN (grant agreement no. 7N5839). S.D. and L.B. acknowledge the 4NanoEARDRM project (EuroNanoMed III co-funded action by the Italian Ministry of University and Research—MIUR). M.M. was supported by the European Union's Horizon 2020 research and innovation program under the Marie Skłodowska-Curie grant agreement COLLHEAR no. 794614. E.M. was supported by the European Union's Horizon 2020 research and innovation program under the Marie Skłodowska-Curie grant agreement HyMedPoly no. 643050.

**Conflicts of Interest:** The authors declare no conflict of interest.